\documentclass[10pt,showkeys,nofootinbib,floatfix,amsmath,amssymb,showpacs,superscriptaddress,notitlepage]{revtex4-1}
\usepackage{hyperref}
\usepackage{caption}
\usepackage{subcaption}
\usepackage{float}
\hypersetup{
colorlinks=true,
urlcolor=blue,
citecolor=blue}
\usepackage[TU]{fontenc}
\usepackage[all]{xy}
\usepackage{amsfonts,amssymb,amsmath,amsgen,amsopn,amsbsy,theorem,graphicx,epsfig}
\usepackage{eufrak,amscd,bezier,latexsym,mathrsfs,enumerate,multirow}
\usepackage{mathtools}
\usepackage{cleveref}
\usepackage[utf8]{inputenc}
\usepackage[english]{babel}
\usepackage[dvipsnames]{xcolor}
\usepackage{academicons}
\definecolor{orcidlogocol}{HTML}{A6CE39}
\usepackage[pagewise]{lineno}
\newcommand{\comment}[1]{}

\begin{document}
\title{$\Lambda_c \to \Lambda$ Form Factors in Lattice QCD}
\author{H. Bahtiyar}
\affiliation{Department of Physics, Faculty of Science and Letters, Mimar Sinan Fine Arts University, Bomonti 34380,
Istanbul, Turkey}
\date{\today}

\begin{abstract}
Studying the semileptonic decays of charmed particles is prominent in testing the standard model of particle physics. Motivated by recent experimental progress in weak decays of the charm baryon sector, we study the form factors of $\Lambda_c \to \Lambda \ell^+ \nu$ transition on two flavor lattices. We compute two- and three-point functions, extract the dimensionless projected correlators, and combine them to form the Weinberg form factors. In the zero transferred momentum limit $f_1$, $f_2$ and $g_1$ form factors are found to be in agreement with other models, furthermore $f_3$ and $g_3$ form factors are comparable to model determinations. The $g_2$ form factor, on the other hand, is found to be mildly larger. We also evaluate the helicity form factors, which is consistent with the previous lattice studies.

\keywords{Charmed baryons, semileptonic form factors, lattice QCD}
\end{abstract}

\maketitle

\section{Introduction}
Recently significant experimental progress has been made regarding the weak decays of heavy baryons. LHCb Collaboration has observed the doubly charmed $\Xi_{cc}^{++}$ particle in the $\Lambda_c^+\,K^-\,\pi^+\,\pi^+$ channel~\cite{Aaij:2017ueg}, Belle Collaboration has measured the branching ratio $\Lambda_c^+ \to p~K^-~\pi^+$~\cite{Zupanc:2013iki} and BESIII has studied the branching fraction of $\Lambda_c^+$ both in the hadronic~\cite{Ablikim:2015flg}~and semileptonic modes~\cite{Ablikim:2015prg,Ablikim:2016vqd,Ablikim:2018woi}. Thanks to the developments in experimental facilities, many decay channels of charmed baryons have been observed, and more precise measurements are expected.

The $\Lambda_c$ baryon is composed of the  $u$, $d$, and $c$ valence quarks, its quantum numbers are $J^P=\frac{1}{2}^+$, and it has a flavor antisymmetric wavefunction. It is a member of the anti-4-plet ($\bar{4}$) in $SU(4)_f$ group formalism. The $\Lambda_c$ baryon is first observed in the 1980s at CERN~\cite{Calicchio:1980sc}. The $\Lambda$ hyperon, on the other hand, is composed of $u$, $d$, and $s$ valence quarks and is well known since the 1960s~\cite{Bhowmik:1963zz}. $\Lambda$ has the spin-parity quantum numbers $J^P=\frac{1}{2}^+$ and a flavor antisymmetric wavefunction. $\Lambda$ and $\Lambda_c$ are members of the exact $SU(4)_f$ anti-4-plet.

Studying the semileptonic decays of charmed particles is prominent in testing the standard model of particle physics. Most of the work focuses on semileptonic decays of charm mesons, yet investigating the decays of charm baryons present additional information. One can gain complementary information about Cabibbo--Kobayashi--Maskawa (CKM) matrix elements and CP violations; moreover, semileptonic charm baryon decays can be used to test the lepton flavor universality~\cite{Faustov:2020thr}. Any deviation from the Standard Model predictions might indicate a hint for the physics beyond the standard model.

Motivated by the theoretical and experimental achievements, we give a timely study of the spin-$\frac{1}{2} \to \text{spin-}\frac{1}{2}$ semileptonic transition of $\Lambda_c \to \Lambda$, which gives access to six form factors, and there is a quark flavor change from charm to strange. In theoretical studies, these six transition form factors play a crucial role in semileptonic decays, which are also used as input for predicting the channels of doubly charmed baryons~\cite{Li:2021qod}. This work is reminiscent of Refs.~\cite{Sasaki:2008ha,Cooke:2014} where hyperon semileptonic form factors have been studied. The simulations have been carried out on $16^3 \times 32$ CP-PACS lattices. We use a ratio of two- and three-point correlation functions and extract the dimensionless projected correlators. The vector and axial-vector form factors are constructed using these correlators. The semileptonic $\Lambda_c \to \Lambda$ decay has been
previously studied in quark models~\cite{Hussain:1990ai,Efimov:1991ex,Cheng:1991sn,Pervin:2005ve,Faustov:2016yza}, QCD sum rules~\cite{MarquesdeCarvalho:1999bqs,Liu:2009sn,Zhao:2020mod}, bag model~\cite{PerezMarcial:1989yh} and lattice QCD~\cite{Meinel:2016dqj}. 

This paper is organized as follows:  Theoretical formalism of the semileptonic decay, details of the lattice setup, and the definition of form factors are given in \Cref{sec:theo}. We present our results, compare them to other works and give a discussion in \Cref{sec:res}. \Cref{sec:con} summarizes our findings.

\section{Theoretical formalism}
\label{sec:theo}
Lattice QCD is a discretized version of QCD. It is an ab initio method that begins directly from the QCD Lagrangian. Calculating QCD numerically on a lattice was introduced nearly forty years ago and has developed into a powerful tool since. Lattice QCD is based on the path-integral representation of quantum field theory, which directly simulates the original theory. Furthermore, form factors can be determined using baryon matrix elements that can be written in terms of QCD path integrals, enabling the lattice gauge theory methods to be used. 

The semileptonic decay $\Lambda_c \to \Lambda$ can be written by combining the vector and the axial-vector currents. The transition matrix element of the process is written in the following form: 
\begin{equation}\label{eq:slff}
\langle \Lambda_c(p') | V_\mu(x) - A_\mu(x) |\Lambda(p) \rangle= \bar{u}_{\Lambda_c}(p') (O_\mu^V(q) - O_\mu^A(q) u_{\Lambda}(p),
\end{equation}
where $V_\mu(x)$ and $A_\mu(x)$ are the vector and the axial-vector currents, respectively, and $q_\mu = p'_\mu - p_\mu$ is the transferred momentum between the incoming and outgoing baryon. The matrix element in \Cref{eq:slff} is parameterized via six form factors which, in the Euclidean space, are given as,
\begin{align}
\label{eq:vecff}
O_\mu^V(q) &= \gamma_\mu f_1(q^2) + \sigma_{\mu \, \nu} q_\nu \frac{f_2(q^2)}{m_{\Lambda_c}+m_{\Lambda}} + i\,q_\mu \frac{f_3(q^2)}{m_{\Lambda_c}+m_{\Lambda}},\\
\label{eq:axff}
O_\mu^A(q) &=\gamma_\mu \gamma_5 g_1(q^2) + \sigma_{\mu \, \nu} q_\nu \gamma_5 \frac{g_2(q^2)}{m_{\Lambda_c}+m_{\Lambda}} + i\,q_\mu \gamma_5 \frac{g_3(q^2)}{m_{\Lambda_c}+m_{\Lambda}},
\end{align}
where the form factors are, the vector $f_1$, the weak magnetism $f_2$, the induced scalar $f_3$, the axial-vector $g_1$, the weak electricity $g_2$ and the induced pseudo-scalar $g_3$. Here $m_{\Lambda_c}$ and $m_{\Lambda}$ denote the mass of $\Lambda_c$ and $\Lambda$ baryons respectively and $\sigma_{\mu \, \nu} = \frac{1}{2i}[\gamma_\mu,\gamma_\nu]$.
$f_1$, $f_2$,  $g_1$ and $g_3$ form factors are called the first-class form factors. Rest of the form factors are called the second-class form factors according to Weinberg’s classification~\cite{Weinberg:1958ut} where the second class form factors transform with the opposite sign under G-parity.

From a phenomenological point of view, form factors are related to the internal structure of the particle at zero transferred momentum. In the flavor symmetric limit, $f_1$ and $f_2$ are related to the electromagnetic form factors while $g_2$ and $f_3$ vanish, so that these form factors are proportional to flavor symmetry breaking at leading order. Using the form factors at zero transferred momentum, one can predict differential decay rates which allow extracting the Cabibbo--Kobayashi--Maskawa matrix element $|V|^2$~\cite{Meinel:2016dqj}.

In order to extract the form factors, we use the correlation functions,
\begin{align}
\label{eq:lambdacorr}
\langle C^{\Lambda\,\Lambda}(t;\textbf{p},\Gamma_4) \rangle &= \sum_x e^{i\textbf{p}x}\, \Gamma_4^{\beta \alpha} \, \langle \text{vac}|T(\chi_\Lambda^\beta(x)\overline{\chi}_\Lambda^\alpha(0))|\text{vac}\rangle, \\
\label{eq:lambdaccorr}
\langle C^{\Lambda_c\,\Lambda_c}(t;\textbf{p},\Gamma_4) \rangle &= \sum_x e^{i\textbf{p}x}\, \Gamma_4^{\beta \alpha} \, \langle\text{vac}|T(\chi_{\Lambda_c}^\beta(x)\overline{\chi}_{\Lambda_c}^\alpha(0))|\text{vac}\rangle, \\
\label{eq:threecorr}
\langle C^{\Lambda_c \mathcal{J}^\mu \Lambda}(t_2,t_1,\textbf{p}^\prime,\textbf{p},\Gamma) \rangle &= - i \sum_{x_1,x_2} e^{-i\textbf{p}x_2}\,e^{i\textbf{q}x_1}\, \Gamma^{\beta\,\alpha} \langle \text{vac}|T(\chi_{\Lambda_c}^\beta(x_2) \mathcal{J}_\mu(x_1)\overline{\chi}_{\Lambda}^\alpha(0))|\text{vac}\rangle,
\end{align}
where \Cref{eq:lambdacorr,eq:lambdaccorr} denote the two-point correlation functions of $\Lambda$ and $\Lambda_c$ baryons, respectively. \Cref{eq:threecorr} is the three-point correlation function, where in this notation a $\Lambda$ baryon is created at time zero, then the external field ($V-A$) is inserted at time $t_1$ and $t_2$ is the time when the $\Lambda_c$ baryon is annihilated. Gamma matrices are defined as $\Gamma_4= \begin{bmatrix}
    1 & 0  \\
    0 & 0 
  \end{bmatrix}$ and $\Gamma_i = \frac{1}{2} \begin{bmatrix}
    \sigma_i & 0  \\
    0 & 0 
  \end{bmatrix}$. 
  The interpolating fields are chosen as 
  \begin{align}
   \chi_{\Lambda} &= \frac{1}{\sqrt{6}} \epsilon_{ijk} \big[2 (u_i^T\,C\gamma_5\,d_j)\,s_k + (u_i^T\,C\gamma_5\,s_j)\,d_k - (d_i^T\,C\gamma_5\,s_j)\,u_k\big], \\
   \chi_{\Lambda_c} &= \frac{1}{\sqrt{6}} \epsilon_{ijk} \big[2 (u_i^T\,C\gamma_5\,d_j)\,c_k + (u_i^T\,C\gamma_5\,c_j)\,d_k - (d_i^T\,C\gamma_5\,c_j)\,u_k\big],   
  \end{align}
where $u$,$d$,$s$ and $c$ denote the individual quark flavors. $C = \gamma_4 \gamma_2$ is the charge conjugation matrix and $i$,$j$,$k$ are the color indicies.

In order to eliminate the time-dependent exponential terms, one needs to define a proper ratio using the two- and three-point functions, 
\begin{align}
	\begin{split}
    R(t_2,t_1,\textbf{p}^\prime,\textbf{p},\Gamma,\mu) = &\frac{\langle C^{\Lambda_c \mathcal{J}^\mu \Lambda}(t_2,t_1,\textbf{p}^\prime,\textbf{p},\Gamma) \rangle}{\langle C^{\Lambda_c \Lambda_c}(t_2;\textbf{p}^\prime;\Gamma_4) \rangle }\\ &\times\Bigg[\frac{\langle C^{\Lambda \Lambda}(t_2-t_1;\textbf{p};\Gamma_4) \rangle \langle C^{\Lambda_c \Lambda_c}(t_1;\textbf{p}^\prime;\Gamma_4) \rangle \langle C^{\Lambda_c \Lambda_c}(t_2;\textbf{p}^\prime;\Gamma_4) \rangle }{\langle C^{\Lambda_c \Lambda_c}(t_2-t_1;\textbf{p}^\prime;\Gamma_4) \rangle \langle C^{\Lambda \Lambda}(t_1;\textbf{p};\Gamma_4) \rangle \langle C^{\Lambda \Lambda}(t_2;\textbf{p};\Gamma_4) \rangle }\Bigg]^{\frac{1}{2}}.
\label{eq:Ratiospinhalf}
	\end{split}
\end{align}
In order to reduce the noise related to the wall-source/sink method~\cite{Can:2012tx}, which is employed for quark smearings, the ratio in~\Cref{eq:Ratiospinhalf} is chosen among the other alternatives~\cite{Bahtiyar:2016dom}. This type of ratio is also preferred in the semileptonic decays of hyperons~\cite{Sasaki:2008ha,Cooke:2014}. In the large Euclidean-time limit time dependence of the correlators cancel, and the ratio is reduced to
  \begin{equation}
     R(t_2,t_1;\textbf{p}^\prime,\textbf{p};\Gamma;\mu)  \xrightarrow[t_1\gg a]{t_2 -t_1 \gg a} \Pi(\textbf{p}^\prime,\textbf{p};\Gamma;\mu).
     \label{eq:RtoPi}
  \end{equation}
We follow the notation given in~\cite{Sasaki:2008ha}. To this end, we first define the dimensionless projected correlators. For the vector part, dimensionless projected correlators are defined as
\begin{align}
\label{eq:veclam0}
 \lambda^V_0 (q^2) &= \sqrt{\frac{2 E_{\Lambda}}{E_\Lambda + m_\Lambda}} \,\Pi(\vec{p},0,\Gamma_4,4),\\
\label{eq:veclams}
 \lambda^V_S (q^2) &= \frac{\sqrt{2 E_{\Lambda}(E_\Lambda + m_\Lambda)}}{p_i} \,\operatorname{Im}(\Pi(\vec{p},0,\Gamma_4,i)),\\
\label{eq:veclamt}
 \lambda^V_T (q^2) &= \frac{\varepsilon_{ijk}\sqrt{2 E_{\Lambda}(E_\Lambda + m_\Lambda)}}{p_j} \,\Pi(\vec{p},0,\Gamma_k,i),
\end{align}
where $i$,$j$,$k$ goes from 1 to 3. Here $\Gamma_4$ and $\mu =4$ are the temporal part of the projection matrix and the Lorentz index of the external current, respectively.

For the axial part of the current, one defines similar dimensionless projected correlators,
\begin{align}
\label{eq:axlam0}
 \lambda^A_0 (q^2) &= \frac{\sqrt{2 E_{\Lambda} (E_\Lambda + m_\Lambda)}}{p_i} \,\Pi(\vec{p},0,\Gamma_i,4),\\
\label{eq:axlaml}
 \lambda^A_L (q^2) &= \sqrt{\frac{2 E_{\Lambda}}{(E_\Lambda + m_\Lambda)}} \,\operatorname{Im}(\Pi(\vec{p},0,\Gamma_i,i)),\\
\label{eq:axlamt}
 \lambda^A_T (q^2) &= \frac{(E_\Lambda+m_\Lambda)(m_\Lambda+m_{\Lambda_c})}{p_i p_j} \,\Pi(\vec{p},0,\Gamma_j,i \neq j).
\end{align}
Note that in \Cref{eq:veclams,eq:axlaml} the imaginary part of the ratio is used. For \Cref{eq:veclam0,eq:veclams,eq:veclamt,eq:axlam0,eq:axlaml} an average over specific combinations of $\mu$ and $\Gamma$ can be taken. \Cref{eq:axlamt} is not computable for some $q^2$ values, so we use a dipole approximation to interpolate and access the $q^2$ values that we are interested in.

We use the two flavor configurations generated by the CP-PACS collaboration~\cite{AliKhan:2001xoi}. These configurations are generated with the $O(a)$-improved Wilson (clover) quark action and a renormalization group improved (Iwasaki) gauge action. We use the clover action for computing the valence quark propagators as well. The hopping parameters of $u$ and $d$ quarks are selected equal to that of light sea quarks $\kappa_{val}^{u,d} = \kappa_{sea}^{u,d}= 0.1410$ which corresponds to $m_\pi \approx 550$ MeV~\cite{Erkol:2008yj}. Details of the configurations are given in~\Cref{table:gauge}.

\begin{table}[!htb]
\caption{The details of the gauge configurations that have been used in this work~\cite{AliKhan:2001xoi}. $N_s$ is the spatial and $N_t$ is the temporal size of the lattice, $a$ is the lattice spacing, $L$ is the spatial extent in physical units, $\beta$ is the inverse gauge coupling, $c_{sw}$ is the clover coefficient and $\kappa_{sea}$ is the hopping parameter of the sea quarks.}
\medskip
\centering
\begin{tabular*}{0.8\textwidth}{@{\extracolsep{\fill}}||l|l|l|l|l|l|l||}
     \hline
 $N_s \times N_t$  &  $a$ (fm) & $L$ (fm) & $\beta$ & $c_{sw}$& $\kappa_{sea}$  & $\#$ of conf. \\
 \hline
 $16^3 \times 32$ & $0.1555$ ($17$)  &  $2.5$ & $1.95$ & $1.530$ & $0.1410$ & $245$ \\
  \hline
\end{tabular*}
\label{table:gauge}
\end{table}

The hopping parameter is chosen as $\kappa^s =0.1393$ for the strange valance quarks, which reproduces the kaon mass~\cite{AliKhan:2001xoi}. Charm quark hopping parameter is tuned to $\kappa^c =0.1045$ using the experimental 1S spin-averaged charmonium mass.

The three-point correlation functions are calculated with a separation of $10$ lattice units in the temporal direction between the source and the sink, corresponding to $t \sim 1.5$ fms. This separation is also preferred in studies of the axial charge~\cite{Sasaki:2003jh} and weak matrix elements~\cite{Sasaki:2007gw} of the nucleon, and semileptonic decays of hyperons~\cite{Sasaki:2008ha}. Since increasing the source-sink separation causes noise in the signal and decreasing the range increases the excited-state contamination, determining the source-sink separation correctly plays an essential role.

 To increase the statistics, we employ positive and negative momenta in all spatial directions and make simultaneous fits over all the data. We also use multiple source-sink pairs, shifting them $10$ lattice units along the temporal direction.
 
 For the vector part of the current, we use the point-split lattice vector current,
  \begin{equation}
		 j_\mu = \frac{1}{2} [\overline{q}(x+\mu) U_\mu^{\dagger}(1+\gamma_\mu)q(x)-\overline{q}(x)U_\mu(1-\gamma_\mu)q(x+\mu)],
		 \label{eq:pointsplit}
  \end{equation}
which is conserved by the Wilson fermions. On the other hand, the axial part of the current needs to be renormalized, for which  we follow the procedure given in~\cite{Sasaki:2008ha} and use 
\begin{equation}
		 Z_A^{c\bar{s}}(m_c,m_s) = \sqrt{Z_A^{c\bar{s}}(m_s,m_s)\,Z_A^{c\bar{s}}(m_c,m_c)},
		 \label{eq:renormax}
  \end{equation}
where the renormalization constants for the $s$ and $c$ quarks are calculated perturbatively~\cite{AliKhan:2001xoi}. We use wall-source/sink method~\cite{Can:2012tx} which is a gauge-dependent object; therefore, we fix the gauge to Coulomb gauge.

We use a modified version of the Chroma software~\cite{Edwards:2004sx} in our computations. The statistical errors are estimated using the single-elimination jackknife method.

\section{Results and discussion}
\label{sec:res}
\begin{figure*}[htb]
    \centering
    \includegraphics{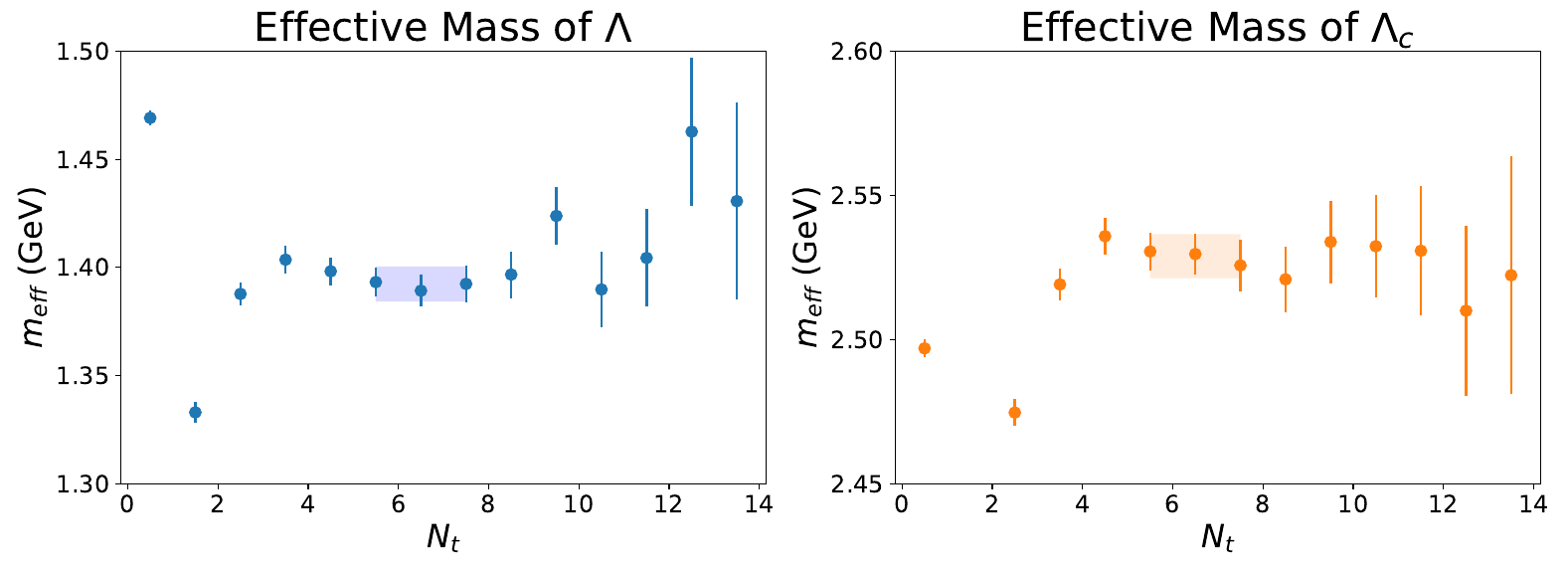}
\caption{Effective mass plots of $\Lambda$ and $\Lambda_c$ particles. Shaded parts represent the fit region and one sigma errors.}
\label{fig:mass}
\end{figure*}

We begin our work with computing the masses of $\Lambda$ and $\Lambda_c$ baryons using the two-point correlation functions given in \Cref{eq:lambdacorr,eq:lambdaccorr}. The two-point correlation function includes all possible states, including the excited states. We perform standard effective mass analysis using the forms given below

\begin{align}
      \label{eq:effmass}
    m_{eff}\bigg(t+\frac{1}{2}\bigg) &= ln \frac{C(t)}{C(t+1)},\\
        \label{eq:expfit}
    C(t) &= Z_\alpha e^{-m t}.
  \end{align}
When the ground state is dominant, signal we obtain from \Cref{eq:effmass} forms a plateau. To this end, we seek a plateau in \Cref{fig:mass} to estimate suitable fit ranges for one-exponential fits. We perform fits using the function given in \Cref{eq:expfit} to extract the ground state masses from the correlation functions.

Masses are found as $m_{\Lambda_c} = 2.529 (7)$ MeV and $m_{\Lambda} = 1.381 (7)$ MeV, which are higher than their experimental values since the light quark masses are unphysically heavy. Since all given errors are statistical errors in this work, a discussion on possible systematic errors is in order. Systematic errors due to unphysically heavy light-quarks could be reduced by repeating the calculations at lighter quark masses and performing an extrapolation to the chiral limit ($m_\pi = 0$), which is planned for future work. Conversely, the form factor results are not too much responsive to the changes in baryon masses~\cite{Bahtiyar:2016dom}. Effects of the light quark mass on the electromagnetic form factors of charmed baryons have been found to be around $10\%$~\cite{Can:2013tna}. Lattice studies on semileptonic decays also report that this effect is negligible~\cite{Aoki:1997np}.

The finite size of the lattice also causes systematic errors. One needs to simulate at several lattices with different volumes to evaluate the finite-volume effects. The configuration sets that have been used in this work have $m_\pi~L \sim 7$, above the rule of thumb bound of $4$, for which the finite-volume effects are considered to be negligible~\cite{Can:2012tx}. Another systematic error comes from the discretization of the lattice. In practice, one should simulate at different lattice spacings and perform an extrapolation to $a \to 0$. However, these calculations are currently beyond our computational resources.  

\begin{figure*}[htb]
    \centering
        \includegraphics{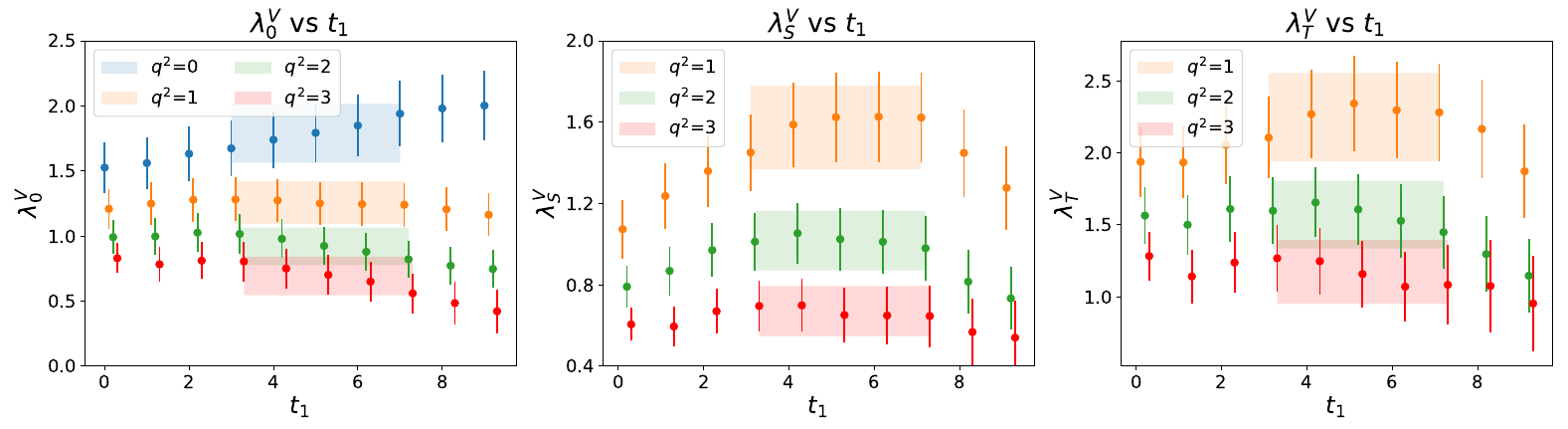}
    \caption{Vector component of the dimensionless correlators ($\lambda^V$) as a function of current insertion time $t_1$. Shaded parts represent the fit region and one sigma errors.}
    \label{fig:vec}
\end{figure*}

\begin{figure*}[htb]
    \centering
        \includegraphics{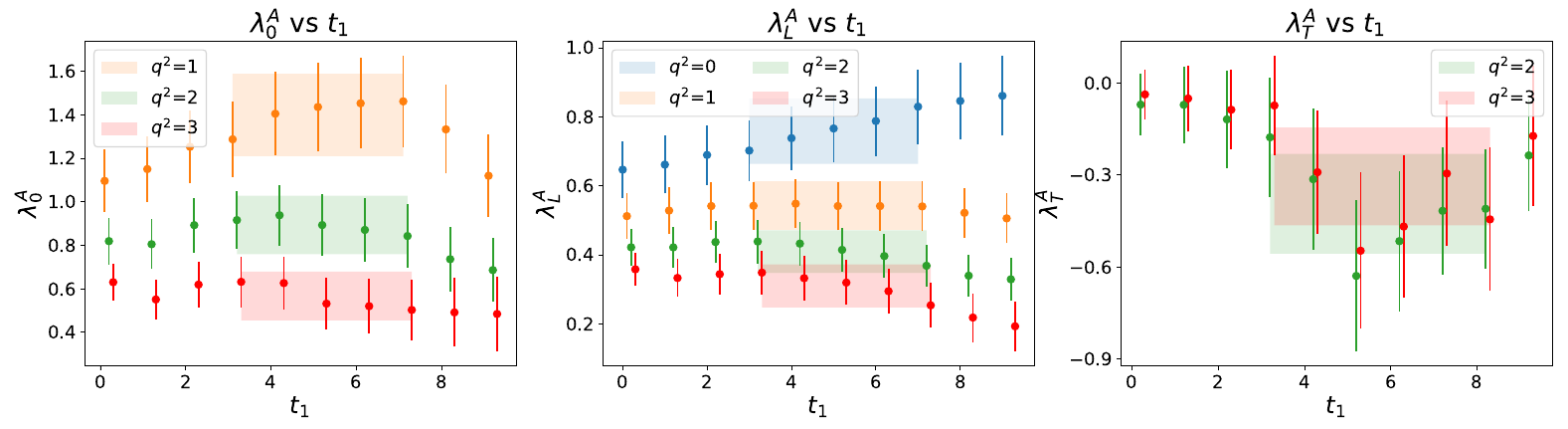}
    \caption{Axial-vector component of the dimensionless correlators ($\lambda^A$) as a function of current insertion time $t_1$. Shaded parts represent the fit region and one sigma errors.}
    \label{fig:ax}
\end{figure*}

We continue our work with calculating the dimensionless correlators given in~\Cref{eq:veclam0,eq:veclams,eq:veclamt,eq:axlam0,eq:axlaml,eq:axlamt}. We plot the correlators as a function of current insertion time for each transferred three-momentum square in~\Cref{fig:vec,fig:ax} and search for plateaus to exclude the excited state contamination. Ground state signals are found in the middle region between the source and sink points. We obtain clean signals both in the vector and axial-vector channels. 

In semileptonic decays, zero three-momentum transfer is also an important parameter, which usually is called as $\textbf{q}^2_{MAX}$ in four momenta. It is clearly seen from~\Cref{fig:vec,fig:ax} that $\vec{q}^2=0$ is accessible for $\lambda^V_0$ and $\lambda^A_L$. To access $\vec{q}^2=0$ for the remaining dimensionless correlators, we employ a dipole extrapolation that we used in our previous work~\cite{Bahtiyar:2016dom}.

Specific combinations of three-point functions are written to access the vector and axial-vector components of the current~\cite{PhysRevD.68.054509}. One extracts each form factor via
\begin{align}
 f_1(\textbf{q}^2) &= \frac{M_{\Lambda_{c}}+M_\Lambda}{2M_{\Lambda_{c}}} 
\left[{\lambda}^{V}_0  - \frac{E_\Lambda-M_{\Lambda_{c}}}{E_\Lambda+M_\Lambda} {\lambda}^{V}_S - \frac{M_\Lambda^2+M_{\Lambda_{c}}^2-2E_\Lambda M_{\Lambda_{c}}}{(M_\Lambda +M_{\Lambda_{c}})(E_\Lambda +M_\Lambda)} {\lambda}^{V}_T
\right] ,\\
 f_2(\textbf{q}^2) &= \frac{M_{\Lambda_{c}}+M_\Lambda}{2M_{\Lambda_{c}}} 
\left[-{\lambda}^{V}_0  + \frac{E_\Lambda-M_{\Lambda_{c}}}{E_\Lambda+M_\Lambda} {\lambda}^{V}_S + \frac{M_\Lambda+M_{\Lambda_{c}}}{E_\Lambda +M_\Lambda} {\lambda}^{V}_T
\right] ,\\
 f_3(\textbf{q}^2) &= \frac{M_{\Lambda_{c}}+M_\Lambda}{2M_{\Lambda_{c}}} 
\left[-{\lambda}^{V}_0  + \frac{E_\Lambda+M_{\Lambda_{c}}}{E_\Lambda+M_\Lambda} {\lambda}^{V}_S + \frac{M_\Lambda-M_{\Lambda_{c}}}{E_\Lambda +M_\Lambda} {\lambda}^{V}_T
\right] ,
\end{align}
for the vector form factors, and
\begin{align}
 g_1(\textbf{q}^2) &= \frac{M_{\Lambda_{c}}+M_\Lambda}{2M_{\Lambda_{c}}} 
\left[{\lambda}^{A}_L  - \frac{M_\Lambda-M_{\Lambda_{c}}}{M_{\Lambda_{c}}+M_\Lambda}\left( {\lambda}^{A}_0 + \frac{E_\Lambda-M_{\Lambda_{c}}}{M_{\Lambda_{c}}} {\lambda}^{A}_T\right)
\right] ,\\
 g_2(\textbf{q}^2) &= \frac{M_{\Lambda_{c}}+M_\Lambda}{2M_{\Lambda_{c}}} 
\left[{\lambda}^{A}_L  - {\lambda}^{A}_0 - \frac{E_\Lambda-M_{\Lambda_{c}}}{M_{\Lambda_{c}}} {\lambda}^{A}_T
\right] ,\\
 g_3(\textbf{q}^2) &= \frac{M_{\Lambda_{c}}+M_\Lambda}{2M_{\Lambda_{c}}} 
\left[{\lambda}^{A}_L  - {\lambda}^{A}_0 - \frac{E_\Lambda+M_{\Lambda_{c}}}{M_{\Lambda_{c}}} {\lambda}^{A}_T
\right] ,\\
\end{align}
for the axial-vector form factors.

\begin{table}[!htb]
\caption{Form factor results for the semileptonic $\Lambda_c \to \Lambda$ transition at $q^2 = 0$ along with a comparison to the nonlattice methods.}
\medskip
\centering
\begin{tabular*}{\textwidth}{@{\extracolsep{\fill}}||l|l|l|l||l|l|l||}
     \hline
  ~  					&  $f_1(0)$ 		& $f_2(0)$ 		&  $f_3(0)$ 		&$g_1(0)$  	&$g_2(0)$		&$g_3(0)$ \\ \hline
 This Work 				&  $0.687~(138) $	& $0.486~(117) $	& $0.164~(80) $	&$0.539~(101)$	&$-0.388~(100)$	&$-0.359~(283)$\\ \hline \hline
Bag M~\cite{PerezMarcial:1989yh} 	&  $0.35 $		&  $0.09 $		& $ 0.25$		& $0.61$ 	&$-0.04$		&$-0.11$\\ \hline
RQM~\cite{Faustov:2016yza}		&$ 1.14 $		&$0.072$		&$0.252$		& $0.517$		&$-0.697$		&$-0.471$\\ \hline
QSR~\cite{Liu:2009sn} 			&$0.665 $		&$0.285$		& --- 			&$0.665 $		&$-0.285 $		& --- \\ \hline
LFCQM~\cite{Zhao:2018zcb} 		&$0.468$		&$0.222$		& ---			&$0.407$		&$-0.035$		&--- \\ \hline
LFCQM~\cite{Geng:2020gjh}		&$0.67 ~(1)$	&$0.76~(2)$		& ---			&$0.59~(1)$		&$-1.59 ~(5)\times 10^{-3}$&---\\ \hline
CQM~\cite{Gutsche:2015rrt} 		&$0.511$		&$0.289$		&$0.014$		&$0.466$		&$-0.025$		&$-0.400$\\ \hline
LFQM~\cite{Li:2021qod}			&$ 0.706$		&$0.362$		&$0.286$		&$0.624$		&$-0.113$ 		&$-0.598$\\ \hline
\hline
\end{tabular*}
\label{table:ff}
\end{table}

Weinberg form factors of the $\Lambda_c \to \Lambda$ semileptonic decay are studied in light-front constituent~\cite{Zhao:2018zcb,Geng:2020gjh} covariant~\cite{Gutsche:2015rrt} and relativistic~\cite{Faustov:2016yza} quark models, in MIT bag model~\cite{PerezMarcial:1989yh} and QCD sum rules~\cite{Liu:2009sn}. Our results for the Weinberg form factors along with a comparison to the other models are given in~\Cref{table:ff}. Any sign difference due to a choice of convention for $f_2$ and $f_3$ form factors are omitted in~\Cref{table:ff}.

Our results for the $f_1$ and $g_1$ form factors agree within errors with most of the model calculations. For the induced scalar $f_3$ and the induced pseudo-scalar $g_3$ form factors, results are still comparable, even though the literature is limited. Our results are larger for the $f_2$ form factor but still agree with the light-front constituent quark model, covariant quark model, and QCD sum rules. On the contrary, the $g_2$ form factor is larger than most of the model determinations, while within the QCD sum rules prediction range.

The second class form factors are expected to be zero in the $SU(3)_f$ symmetric limit, but since this symmetry is broken in nature, small but a nonzero $g_2$ value is expected for hyperon decays~\cite{Sasaki:2016zpr}. On the other hand, our work is focused on the charm sector, where quark flavor changes from charm to strange. It is well-known that $SU(4)_f$ symmetry is badly broken, thus, we expect to find nonzero second class form factors, which cannot be ignored in observable predictions~\cite{Kubodera:1984qd,Carson:1985pi}. Moreover, since the present work has some limitations e.g., no extrapolation to the chiral limit, small volume $16^3 \times 32$ size lattices, the systematic effects might be significant. 

\begin{figure*}[t!]
    \centering
    \includegraphics{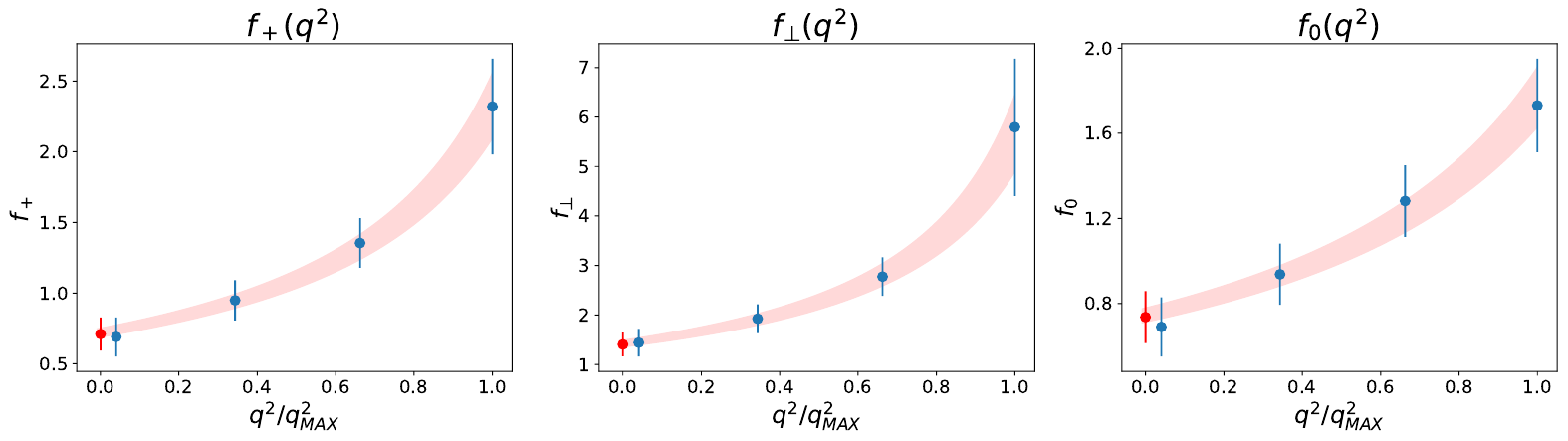}
    \caption{Vector form factors in the helicity-based definition. Blue dots are the extracted numerical results; red shaded bands are the fits to the function given in \Cref{eq:fitfunc}. Red dots are the values obtained from fits.}
    \label{fig:vechel}
\end{figure*}

\begin{figure*}[t!]
    \centering
    \includegraphics{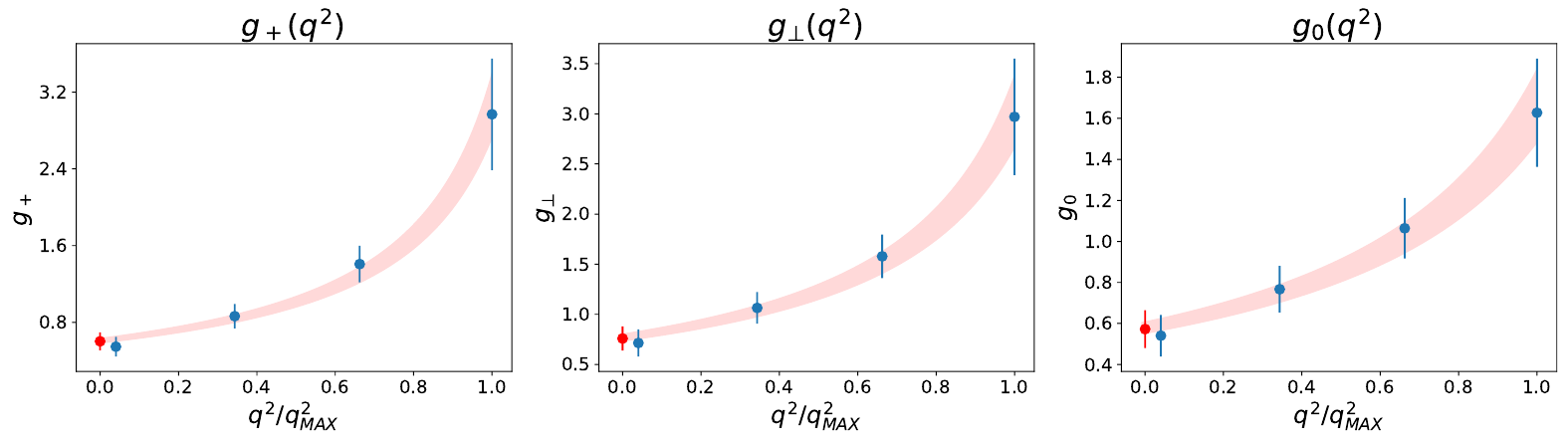}
    \caption{Same as \Cref{fig:vechel} but for axial form factors.}
    \label{fig:axhel}
\end{figure*}

Next step is to determine the helicity-based definition of the form factors to compare our results with another lattice study of Ref.~\cite{Meinel:2016dqj}. Helicity form factors are related to the Weinberg form factors as follows~\cite{Detmold:2015aaa}:
\begin{align}
 f_+(\textbf{q}^2)     &= f_1(\textbf{q}^2) + \frac{\textbf{q}^2}{m_{\Lambda_c}(m_{\Lambda_c}+m_\Lambda)} f_2(\textbf{q}^2), \\
 f_\perp(\textbf{q}^2) &= f_1(\textbf{q}^2) + \frac{m_{\Lambda_c}+m_\Lambda}{m_{\Lambda_c}} f_2(\textbf{q}^2),  \\
 f_0(\textbf{q}^2)     &= f_1(\textbf{q}^2) + \frac{\textbf{q}^2}{m_{\Lambda_c}(m_{\Lambda_c}-m_\Lambda)} f_3(\textbf{q}^2), \\
 g_+(\textbf{q}^2)     &= g_1(\textbf{q}^2) - \frac{\textbf{q}^2}{m_{\Lambda_c}(m_{\Lambda_c}-m_\Lambda)} g_2(\textbf{q}^2), \\
 g_\perp(\textbf{q}^2) &= g_1(\textbf{q}^2) - \frac{m_{\Lambda_c}-m_\Lambda}{m_{\Lambda_c}} g_2(\textbf{q}^2), \\
 g_0(\textbf{q}^2)     &= g_1(\textbf{q}^2) - \frac{\textbf{q}^2}{m_{\Lambda_c}(m_{\Lambda_c}+m_\Lambda)} g_3(\textbf{q}^2). \\
\end{align}
 The equations maintain the endpoint relations for the helicity form factors \cite{Hiller:2021zth}. We extract the helicity form factors from our Weinberg form factors and extrapolate to zero momentum via
\begin{equation}
f/g(\textbf{q}^2)=\frac{f/g(0)}{1-\frac{\textbf{q}^2}{m^2}},
 \label{eq:fitfunc}
\end{equation}
where $f/g(0)$ and $m$ are the free fit parameters. Here $f/g(0)$ represents the form factor result at zero transferred momentum. This type of function is also used in QCD sum rules analyses~\cite{Azizi:2011mw} to describe the form factor behavior and preferred for the weak form factors of $c$ quark decays~\cite{Hu:2020mxk}. As a check, we employed different types of fit functions containing higher-order momentum terms, as used in Refs.~\cite{Geng:2020gjh,Li:2021qod,Liu:2009sn}. We find our zero momentum results deviate by $5\%-10\%$ only. We plot the helicity-based form factors in~\Cref{fig:vechel,fig:axhel}. Our results are mildly larger than the other lattice calculation of Ref.~\cite{Meinel:2016dqj}. This is most probably due to the different systematics of the lattice setups. Besides, results of Ref.~\cite{Meinel:2016dqj} are obtained at the chiral limit with the z-expansion fit procedure. We could not apply this procedure to our data since our results are not extrapolated to the chiral limit. Nevertheless, the qualitative properties of the form factors are similar to those obtained in Ref.~\cite{Meinel:2016dqj}.

Finally, the decay width and other physical quantities can be calculated using the form factors extracted in this work. However, we note that the systematic effects must be controlled better and accounted for before making any phenomenologically relevant predictions.

\section{Conclusion}
\label{sec:con}
We studied the semileptonic decay of $\Lambda_c \to \Lambda \bar{\ell} \nu_\ell$ in a $2$-flavor $16^3 \times 32$ lattice QCD simulation. We have extracted the dimensionless projected correlators, which lead to the Weinberg form factors. We have also computed the helicity form factors and compared our results to the other available lattice study. Most of our form factor results qualitatively agree with the predictions of other nonlattice works. However, there seems to be a quantitative disagreement for the $g_2$ form factor, which calls for more investigations to resolve.

Studying the semileptonic decays of charmed particles play a critical role in understanding the weak and strong interactions. Charm quark decays are also crucial for understanding the Standard Model parameters and searching for the signs beyond the standard model physics. Besides, examining the form factors of charm baryons present valuable information on their internal structure. The semileptonic transition of charm baryons have been studied in quark models~\cite{Hussain:1990ai,Efimov:1991ex,Cheng:1991sn,Pervin:2005ve,Faustov:2016yza}, QCD sum rules~\cite{MarquesdeCarvalho:1999bqs,Liu:2009sn,Zhao:2020mod,Azizi:2011mw}, bag model~\cite{PerezMarcial:1989yh} and lattice QCD~\cite{Meinel:2016dqj,Meinel:2021rbm,Zhang:2021oja}.

We have recently implemented a relativistic heavy quark action to our framework and studied the charm baryon spectrum~\cite{Bahtiyar:2020uuj} and electromagnetic transition form factors of doubly charmed baryons~\cite{Bahtiyar:2018vub}. This work serves as a benchmark before extending our studies to the semileptonic form factors of charmed and bottomed hadrons.

\section*{Acknowledgment}
The unquenched gauge configurations used in this work were generated by CP-PACS collaboration~\cite{AliKhan:2001xoi}. We used a modified version of the Chroma software system~\cite{Edwards:2004sx}. The author thanks K.U. Can for valuable discussions and his comments on the manuscript, R. Zwicky for discussions on endpoint relations and T. T. Takahashi for sharing the gauge configurations. The publicly available configurations are downloaded via the ILDG/JLDG network~\cite{Amagasa:2015zwb}. The numerical calculations reported in this paper were partially performed at TÜBİTAK ULAKBİM, High Performance and Grid Computing Center (TRUBA resources).

\bibliographystyle{elsarticle-num}
\bibliography{semi.bib}

\end{document}